\documentclass[twocolumn]{jpsj3}
\usepackage{graphicx}
\usepackage{dcolumn}
\usepackage{bm}

\title{Signature of a  $Z_2$ vortex in the dynamical correlations of the triangular-lattice Heisenberg antiferromagnet}

\author{Tsuyoshi Okubo\thanks{E-mail address: okubo@spin.ess.sci.osaka-u.ac.jp}, Hikaru Kawamura 
}
\inst{Department of Earth and Space Science, Faculty of Science, Osaka University, Toyonaka 560-0043, Japan 
}
\abst{Dynamical properties of the classical Heisenberg antiferromagnet on the triangular lattice are investigated by means of a spin-dynamics simulation and an analytical calculation. While the model was suggested to exhibit a topological transition driven by the $Z_2$-vortex binding-unbinding, weakness of the associated thermodynamic singularity has made it difficult to observe the evidence of the $Z_2$ vortex experimentally so far, only some indirect support. Here, we show that the signature of the $Z_2$-vortex excitation and  the vortex-induced topological transition can be captured from the dynamical spin correlations. In particular, the dynamical spin structure factor exhibits a characteristic central peak around the $Z_2$-vortex transition, and this central peak will be a fingerprint of the $Z_2$-vortex excitation.}

\kword{frustration, triangular lattice, dynamical spin structure factor, $Z_2$ vortex, spin-dynamics simulation, topological transition}

\begin{document}
\maketitle
\section{Introduction}
 Recently, geometrically frustrated magnets have attracted much
 interest because of their unconventional ordering behaviors, including
 the possible quantum spin-liquid state \cite{Anderson} and the novel
 ordered states like chiral ordered state \cite{MiyashitaShiba,Onoda},
 spin  nematic or multipolar state \cite{Zhitomirsky,Tsunetsugu},
 spin-gel state \cite{KawamuraYamamoto2} {\it etc\/}. Such novel ordered
 states as well as the associated phase transitions are often borne by
 novel excitations inherent to frustrated systems. One typical example
 might be a $Z_2$ vortex, stabilized in a class of two-dimensional (2D)
 Heisenberg magnet with the locally noncollinear spin order
 \cite{KawamuraMiyashita}, {\it e.g.\/}, the frustrated
 Heisenberg antiferromagnet (AF) on the triangular lattice (Fig.1).

 Indeed, the AF Heisenberg model on the 2D triangular lattice is a typical example of geometrically frustrated magnets and has been extensively studied as the standard model of frustrated systems. In recent experiments on the triangular-lattice  Heisenberg AFs $S$=3/2 NaCrO$_2$  \cite{Olariu,Hsieh,Hsieh2,Hemmida}, $S$=1 NiGa$_2$S$_4$  \cite{Nakatsuji,Nambu,Takeya,Yaouanc,MacLaughlin,Yamaguchi,Yamaguchi2} and $S$=1/2  organic compounds, $\kappa$-(BEDT-TTF)$_2$Cu$_2$(CN)$_3$ \cite{Shimizu,Yamashita,Yamashita2} or EtMe$_3$Sb[Pd(dmit)$_2$]$_2$ \cite{Tamura,Itou}, a spin-liquid like  behavior without the standard magnetic long-range order was observed, while all of these compounds  exhibit a weak but clear transition-like anomaly at a  finite temperature.  As a possible explanation of the observed experimental behavior, ref. 6 suggested a $Z_2$ vortex as a key ingredient. Experimentally observed weak anomaly was then ascribed to the $Z_2$-vortex driven topological transition, while the observed low-temperature state to the ``spin-gel'' state where the spin correlation length is kept finite while the ergodicity is broken topologically \cite{KawamuraYamamoto2,KawamuraYamamoto1}.

 If one considers the simplest case of the classical AF Heisenberg model
 with the nearest-neighbor AF coupling, the ground state
 is the so-called 120-degrees structure. It was demonstrated by Kawamura
 and Miyashita that the model sustained a topologically stable point
 defect characterized by a two-valued topological quantum number, a
 $Z_2$ vortex \cite{KawamuraMiyashita} (Fig.1). They suggested that,
 though the AF spin correlation length remains finite at any finite
 temperature $T>0$, a thermodynamic phase transition driven by the
 binding-unbinding of the $Z_2$ vortices occurred at a finite
 temperature.
\begin{figure}
 \begin{center}
 \includegraphics[]{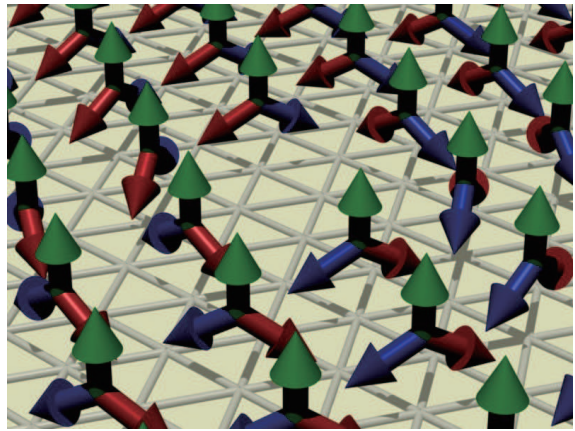}
  \end{center}
 \caption{ (Color online) Illustration of a  $Z_2$ vortex formed by the local 120-degrees
 structure realized in the antiferromagnetic Heisenberg model on the
 triangular lattice.}
\end{figure}

 The remarkable feature of the $Z_2$-vortex transition might be the
 decoupling of the two length scales, a spin-wave (SW) correlation
 length $\xi_{sw}$ and a vortex correlation length $\xi_v$, the latter
 corresponding to the mean separation of free $Z_2$ vortices
 \cite{KawamuraYamamoto2}. At the $Z_2$-vortex transition temperature
 $T=T_v$, the vortex correlation length $\xi_v$ diverges, while the
 spinwave correlation length $\xi_{sw}$ is kept finite (Fig.2). In the
 high-temperature regime $T \gg T_v$ where $\xi_v \ll \xi_{sw}$, the
 standard  spin correlation length $\xi$ is dominated by vortices $\xi
 \simeq \xi_v$, while, in the low-temperature regime, $\xi$ is dominated
 by spinwaves $\xi \simeq \xi_{sw}$. Note that, because of the
 finiteness of $\xi_{sw}$,  the full spin correlation length is kept
 finite even at $T < T_v$. The relative importance of vortices and
 spinwaves is interchanged around a crossover temperature $T_\times
 (>T_v)$ at which $\xi_v\simeq \xi_{sw}$ (Fig.2). Such a decoupling
 behaviour is in sharp contrast to the behavior in other vortex-driven
 transitions, {\it i.e.\/}, the Kosterlitz-Thouless (KT) transition of
 the 2D {\it XY\/} model, where there exists only one length scale and
 the spin correlation length stays infinite throughout the low
 temperature phase \cite{KT1,KT2}.

\begin{figure}
 \begin{center}
 \includegraphics[]{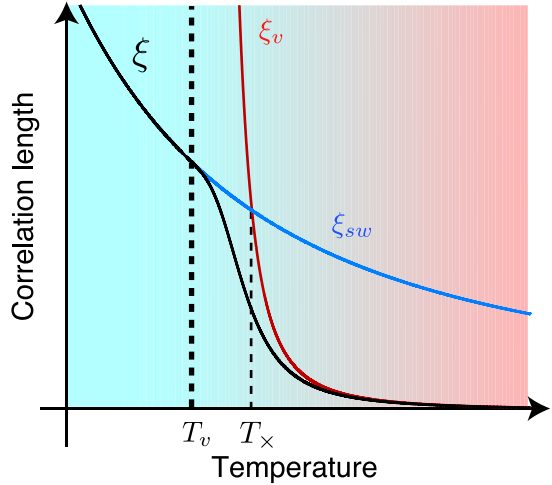}
  \end{center}
 \caption{(Color online) Schematic picture of the decoupling between vortices and spinwaves, each characterized by the associated correlation length, $\xi_v$ and $\xi_{sw}$. At the $Z_2$-vortex  transition temperature $T_v$, the vortex correlation-length $\xi_v$ diverges,  while the spinwave correlation length $\xi_{sw}$ is kept finite. The behavior of the full spin correlation length $\xi$ is also shown. In  the low-temperature spin-gel state, the spin correlation is dominated by spinwave excitations, whereas, in the high-temperature phase, the spin correlation is dominated  by vortex excitations. There exists a crossover temperature $T_\times (>T_v)$ at which the relative importance of vortices and spinwaves are interchanged. $T_v$ and $T_\times$ were estimated by the recent Monte Carlo simulation to be $T_v = 0.285\pm 0.005$  and $T_\times\simeq 0.294$ (ref. 6). }
\end{figure}

According to a recent theoretical analysis \cite{KawamuraYamamoto2}, the diverging behavior of $\xi_{v}$ is given by
\begin{equation}
 \xi_v \sim \exp \left[\left(\frac{A}{T-T_v}\right)^\alpha\right],\ \ \ T >T_v,
\end{equation}
where $\alpha < 1$. The usual thermodynamic macroscopic observables such as the specific heat or the magnetic susceptibility show only a week essential singularity at $T_v$, which makes the direct experimental observation of the $Z_2$-vortex transition rather difficult. So far, only an indirect experimental support was reported \cite{MacLaughlin,Yamaguchi,Olariu}.

 In this paper, we concentrate on the {\it dynamical\/} properties of
 the model in search for the signature of a $Z_2$ vortex. Our numerical
 result indicates that the signature of a $Z_2$-vortex and a
 $Z_2$-vortex driven topological transition can be captured from the
 information of {\it dynamical} spin correlations, specifically from
 that of the dynamical spin structure factor, which can be measured by
 means of inelastic neutron scattering.
 
The rest of the paper is organized as follows. In Sec. \ref{Method}, we describe the model and details of our spin-dynamics simulation. In Sec. \ref{Results}, we present the result of our numerical simulation and analytical calculation, and discuss their relation to the $Z_2$-vortex excitation.  In Sec. \ref{Conclusion}, we conclude with a summary and implications to experiments.

\section{Model and Method}
\label{Method}
 The model we consider is the classical Heisenberg AF on the two-dimensional triangular lattice, whose Hamiltonian is given by 
\begin{equation}
\mathcal{H} = J \sum_{\langle i,j\rangle} \vec{S}_i\cdot\vec{S}_j, 
\end{equation}
where $J > 0$, and the sum is taken over all nearest-neighbor pairs. We
perform here a spin-dynamics simulation of the model. The lattice is a
$L\times L$ triangular lattice of a rhomboidal shape with periodic
boundary conditions. The dynamics is assumed to obey the classical
analogue of the Bloch equation, 

\begin{equation}
 \hbar \frac{d \vec{S}_i}{d t} = J\left(\sum_{\delta}\vec{S}_{\delta}\right) \times \vec{S}_i,\label{SpinDynamics_eq}
\end{equation}
where the sum is taken over all nearest neighbours. Note that this dynamics conserves the total energy as well as the total uniform magnetization $\vec M=\sum_i \vec S_i$. The temperature effect is taken into account via initial spin configurations, which are generated from the standard equilibrium Monte Carlo (MC) simulation at the temperature $T$. Dynamical observables are calculated by averaging over $500-1000$ independent spin-dynamics simulations starting from different initial configurations prepared by equilibrium MC runs.

In order to prepare initial spin configurations for our spin-dynamics simulations, we generate equilibrium spin configurations at a given temperature $T$ by first performing a hybrid Monte Carlo simulation which consists of the heat-bath updating and the over-relaxation updating. One hybrid MC step (MCS) consists of one heat-bath and ten over-relaxation sweeps.  At each temperature, we generate such equilibrium spin configurations, which are taken from our MC run every $10^3$ MCS after discarding $10^5$ MCS for equilibration.

 In our spin-dynamics simulation, we employ a recently developed
 second-order decomposition method to solve equation
 \eqref{SpinDynamics_eq} numerically
 \cite{decomposition1,decomposition2}. We set a time step $\delta t
 =0.01\hbar J^{-1}$, and the integration of equation
 \eqref{SpinDynamics_eq} is carried out to $t_{max}$, typically of the
 order of $t_{max} = 800\hbar J^{-1}$. We checked the accuracy of our
 simulation by comparing the obtained data at several temperatures with those obtained with a finer time step of $\delta t=0.001\hbar J^{-1}$, and with those obtained by using the fourth-order Runge-Kutta method with a time step of $\delta t=0.01\hbar J^{-1}$.

 The dynamical spin structure factor is defined by

\begin{equation}
S(\bm{q}, \omega) \equiv \int d\bm{r}\int dt \left \langle
                                          \vec{S}_{\bm{0}}(0)\cdot
                              \vec{S}_{\bm{r}}(t)\right\rangle
                              \exp[-i(\omega t+\bm{q}\cdot\bm{r})].
\end{equation}
where $\omega$ is the angular frequency, $\bm{q}$ is the wavevector, and $\left \langle\cdots \right\rangle$ denotes a thermal average.

 The dynamical spin structure factor is calculated via the formula,
\begin{equation}
 \tilde{S}(\bm{q},\omega) = \left\langle |\vec{S}_{\bm{q}}(\omega)|^2 \right\rangle,
\end{equation}
where $\langle \cdots \rangle$ represents an average over initial spin configurations, and $\vec{S}_{\bm{q}}(\omega)$ is the space-time Fourier transform of the spin variable given by
\begin{equation}
 \vec{S}_{\bm{q}}(\omega) \equiv
 \int d t\sum_{i}\vec{S}_i(t)\exp[-i(\bm{q}\cdot\bm{r}_i+\omega t) ] .
\end{equation}
In order to decrease statistical errors, the dynamical spin structure
factor is symmetrized as
\begin{equation}
 S(\bm{q},\omega) \equiv  \frac{\tilde{S}(\bm{q},\omega)+
                                 \tilde{S}(\bm{q},-\omega)}{2}.
\end{equation}

The central component of $S(\bm{q},\omega)$ is often subject to
pronounced finite-size effect. We have checked the system-size
dependence of $ S(\bm{q},\omega)$ by comparing those of $L=192, 384,
768$ at each temperature for the triangular case, to confirm that the
data shown here are close enough to the bulk ones.

\section{Results}
\label{Results}
 In Fig.3a, we show the frequency dependence of the dynamical spin
 structure factor $S(\bm{q}, \omega)$ at a wavevector $\bm{q}=(4\pi/3 +
 4\pi/192, 0)$, slightly off the K-point. Note that the perfect
 120-degrees structure yields the magnetic Bragg peak at the K-point,
 $\bm{q}_{K} = (4\pi/3, 0)$.  The temperature range studied is below
 around the specific-heat peak temperature $T_{peak}\simeq 0.32$,
 including both above and below $T_v\simeq 0.285$. As can be seen from
 the figure, at a temperature $T=0.280$ slightly below $T_v$,
 $S(\bm{q},\omega)$ exhibits only two side peaks corresponding to
 spinwave excitations. On increasing the temperature, in addition to the
 spinwave peaks, a {\it central
 peak} centered at $\omega=0$ appears and grows.
 The central peak and the spinwave peaks coexist in the
 range $0.290 \lesssim T  \lesssim 0.305$, whereas, at higher temperatures, these peaks
 are merged into a single broad peak at $\omega  =0$ with spinwave
 shoulders remaining at finite $\omega$.  At $T=0.305$, the intensity of the central  peak is about $1\%$ of that at the K point.  The width of the central peak is of the same amount as that of the spinwave peak, showing no pronounced tendency to sharpen toward $T\rightarrow T_v$. Note that the onset of the central component in $S(\bm{q},\omega)$ is fully correlated to the vortex transition at $T=T_v\simeq 0.285$. This correspondence suggests that the observed central peak might be originated from the contribution of free $Z_2$ vortices.

\begin{figure*}
 \begin{center}
  \includegraphics[width=16cm]{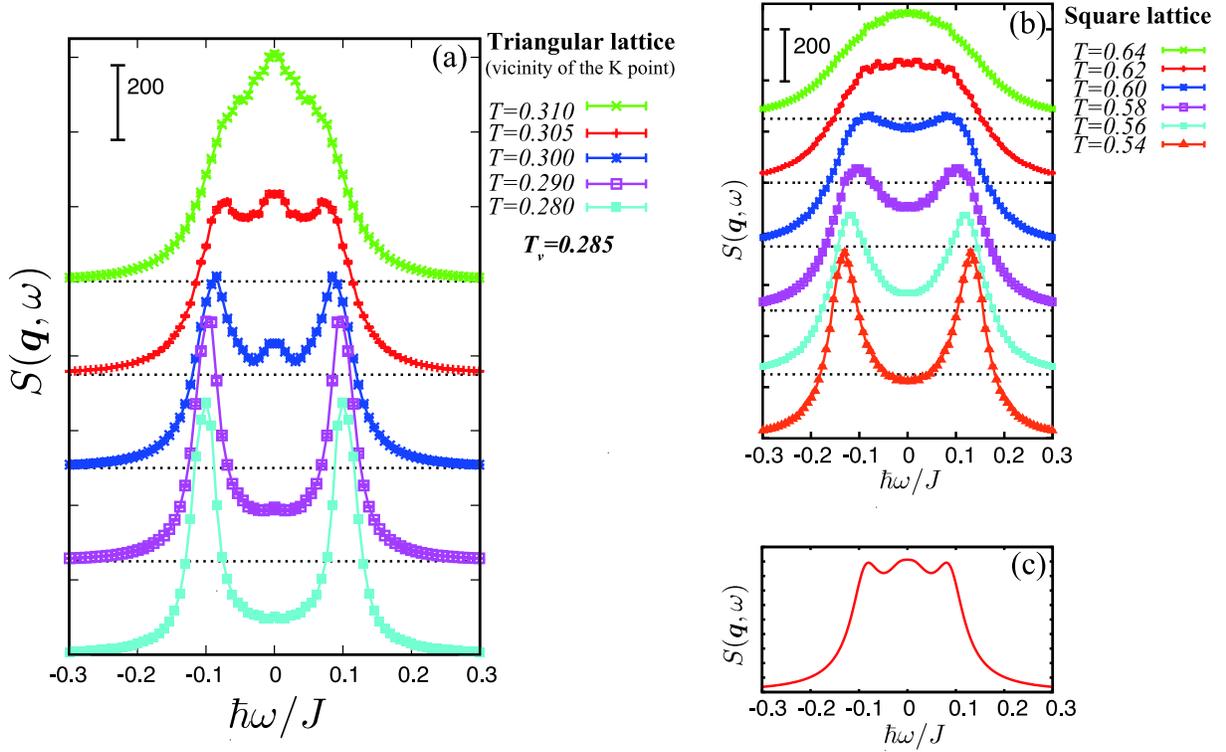}
 \end{center}
 \caption{(Color online) (a) The frequency dependence of the dynamical structure factor of the
 triangular-lattice Heisenberg antiferromagnet at a wavevecotr
 $\bm{q}=(4\pi/3 + 4\pi/192, 0)$ near the K point for the temperatures
 below around the specific-heat maximum temperature $T< T_{peak}\simeq
 0.32$ including both below and above $T_v\simeq 0.285$. The lattice
 size is $L=768$. For better visualization, baselines of the data at
 different temperatures are shifted vertically (dotted lines). (b)
 The frequency dependence of the dynamical structure factor of the
 square-lattice Heisenberg antiferromagnet at a wavevecotr $\bm{q}=(\pi
 - 2\pi/128, \pi-2\pi/128)$ for the temperatures below around the specific-heat
 maximum temperature $T < T_{peak}\simeq 0.68$. The lattice size is
 $L=256$. (c) A typical pattern of the dynamical structure factor
 calculated from eq. \eqref{sqw_ana} by assuming Lorenzian forms
 for $S_{sw}(\bm{q})$ and $S_v(\bm{q})$,  $S_{sw}(\bm{q})\propto
 1/[1+\xi_{sw}^2(\bm{q}-\bm{q}_K)^2]$ and  $S_v(\bm{q}) \propto
 1/[1+\xi_{v}^2\bm{q}^2]$, with a linear dispersion relation,
 $\omega_{sw}(\bm{q})=v_{sw}|\bm{q}-\bm{q}_K|$ and neglecting the $\bm{q}$-dependence of $\tau(\bm{q})$. The parameters in eq. 
 \eqref{sqw_ana} are $\xi_v=250, \xi_{sw}=150$, $\bm{q}=(4\pi/3 + 4\pi/192, 0)$,  $\tau(\bm{q})=40$ and $v_{sw}=1.36$. The central peak arises when $\tau$ is moderately small, $1/\omega_{sw}(\bm{q})  \lesssim \tau \lesssim \xi_{sw}/v_{sw}$.}
\end{figure*}

 In order to further clarify the situation, we also compute
 $S(\bm{q},\omega)$ for the unfrustrated 2D Heisenberg AF
 on a square lattice, a model which does not bear a $Z_2$ vortex. The
 result is shown in Fig.3b for the temperatures below around the specific-heat peak temperature $T_{peak}\simeq 0.68$. In sharp contrast to the triangular case, no central peak is observed in the low-temperature regime where spinwave peaks exist. This absence of the central peak coexisting with the spinwave peaks supports the view that the central peak observed in the triangular case is due to free $Z_2$ vortices.

 We next perform a simplified theoretical analysis of the dynamical spin structure factor based on the spinwave-vortex decoupling picture. We start with a simple assumption of the factorization of the dynamical spin-spin correlation function $C(\bm{r}, t)$ into the vortex part and the spinwave part

\begin{equation}
 C(\bm{r}, t) \equiv \left\langle \vec{S}_{\bm{0}}(0)\cdot \vec{S}_{\bm{r}}(t)\right\rangle \simeq C_{v}(\bm{r},t)C_{sw}(\bm{r},t),
\label{eq_fac}
\end{equation}
where $C_{sw}(\bm{r},t)$ is a dynamical  spin correlation function due to the spinwave excitations and $C_v(\bm{r},t)$ is the one due to the $Z_2$-vortex excitations. We assume that the space Fourier transform of $C_{sw}(\bm{r},t)$ is governed by the spinwave mode with a dispersion $\omega_{sw}(\bm{q})$ and with a damping $\tau_{sw}(\bm{q})$ as

\begin{equation}
 C_{sw}(\bm{q},t) =
 S_{sw}(\bm{q})\exp[-t/\tau_{sw}(\bm{q})\pm i\omega_{sw}(\bm{q})t],
\label{eq_fac_sw}
\end{equation}
where $S_{sw}(\bm{q})$ is the static spin structure factor due to spinwaves. We also assume that the corresponding vortex-part is given by a simple exponential form as

\begin{equation}
 C_{v}(\bm{q},t) = S_v(\bm{q})\exp[-t/\tau_{v}],
\label{eq_fac_v}
\end{equation}
where $S_{v}(\bm{q})$ is the static spin structure factor due to
vortices, and $\tau_{v}$ is assumed to be $\bm{q}$-independent. From
eqs. (8)-(10), $S(\bm{q},\omega)$ is obtained as the convolution of the spinwave part and the vortex part as

\begin{equation}
 S(\bm{q},\omega) = \int d\bm{q}^\prime
 \frac{2\tau(\bm{q}-\bm{q}^\prime)
 S_{sw}(\bm{q}-\bm{q}^\prime)S_v(\bm{q}^\prime)}{1+[\tau(\bm{q}-\bm{q}^\prime)]^2[\omega\pm\omega_{sw}(\bm{q}-\bm{q}^\prime)]^2},
\label{sqw_ana}
\end{equation}
where the reduced correlation time is given by $1/\tau (\bm{q}) \equiv 1/\tau_{sw} (\bm{q})+1/\tau_{v}$. The peak of $S_{sw}(\bm{q})$ is located at the K point, while $S_{v}(\bm{q})$ is expected to be peaked at $\bm{q}=0$. Thus, $S(\bm{q}, \omega)$ typically exhibit two peaks, one at $\omega=\omega_{sw}(\bm{q})$ and the other at $\omega=\omega_{sw}(\bm{q}_K)$. Since $\omega_{sw}(\bm{q}_K) = 0$, the latter corresponds to the central peak. At the $Z_2$-vortex transition $T=T_v$, the vortex correlation length $\xi_v$ diverges, and in the low temperature spin-gel state, $S_v(\bm{q}) \propto \delta(\bm{q})$. Thus, in the spin-gel state below $T_v$, $S(\bm{q},\omega)$ is determined solely by spinwaves without a central-peak structure.
\begin{figure*}
 \begin{center}
  \includegraphics[width=16cm]{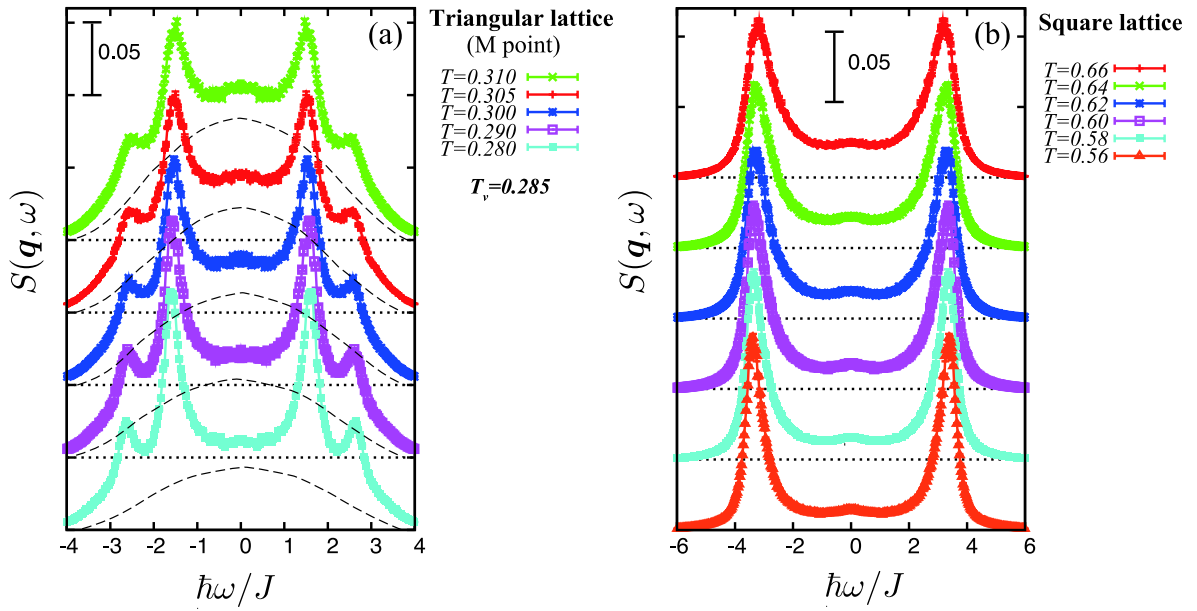}
  \end{center}
\caption{(Color online) (a) The frequency dependence of the dynamical spin structure
 factor of the triangular-lattice Heisenberg antiferromagnet at a
 wavevecotr $\bm{q}=(0, 2\pi/\sqrt{3})$ corresponding to the M point for
 the temperatures as in Fig.3a. The lattice size is $L=768$. For better
 visualization, baselines of the data at different temperatures are
 shifted vertically (dotted lines). As a guide to the eye, the central
 component referred in the main text is indicated by the dashed
 lines. (b) The frequency dependence of the dynamical structure
 factor of the square-lattice Heisenberg antiferromagnet at a wavevecotr
 $\bm{q}=(0, \pi)$ for the temperatures as in Fig.3b. The lattice size is $L=256$. }
\end{figure*}

 In Fig.3c, we show a typical form of $S(\bm{q},\omega)$ above $T_v$
 calculated from eq. \eqref{sqw_ana} under several assumptions given in the
 caption of Fig.3. As can be seen from
 the figure, the observed shape of the dynamical spin structure factor
 is well reproduced by \eqref{sqw_ana},
 at least qualitatively, as long as $\tau(\bm{q})$ is taken to be moderately small. This observation gives further support to the
 view that the central peak of $S(\bm{q},\omega)$ is originated from
 free $Z_2$ vortices.

 Another explanation of the central peak  of $S(\bm{q},\omega)$ was
 suggested by Mertens {\it et al} in terms of the KT vortices of the 2D
 {\it XY\/} model \cite{Mertens}. Based on a dilute vortex-gas picture,
 these authors estimated the free-vortex contribution to
 $S(\bm{q},\omega)$ in the form of squared Lorenzian
 \cite{Mertens}. The data of several spin-dynamics simulations on the 2D {\it XY\/} model were interpreted in terms of this theory \cite{Mertens,DPLandau1,DPLandau2}. In our observation, however, the motion of $Z_2$ vortices are incompatible with such a dilute gas picture. The $Z_2$ vortices of the present model behave  {\it diffusively\/} so that the root-mean-squared displacement of the free vortices is proportional to the square root of the elapsed time $t$ as $\sqrt{\langle \delta r^2(t) \rangle} \propto \sqrt{t}$. Hence, the origin of the central peak observed in the present model probably differs from  the one proposed by Mertens {\it et al} for the KT vortices.

 The central peak of  $S(\bm{q},\omega)$ is not limited to the vicinity
 of the K point. A broad central component also appears even at
 wavevectors away from the K point, though its intensity is
 significantly reduced and its width is much broadened. In Fig.4a, we show the $\omega$-dependence of $S(\bm{q},\omega)$ at the M point, $\bm{q}=\bm{q}_M= (0, 2\pi/\sqrt{3})$. In addition to the spinwave side peaks, a broad but distinct central component  centered at  $\omega =0$ exists. Note that the width of the central component is larger than that near the K point by an order of magnitude, while its intensity is weaker than that near the K point by a factor of $10^{-3}$. The intensity of the central component tends to be decreased with decreasing the temperature as in the vicinity of the K point.

 For comparison, we show in Fig.4b $S(\bm{q},\omega)$ of the
 square-lattice Heisenberg AF at the zone boundary,
 $\bm{q}=(0,\pi)$. The computed $S(\bm{q},\omega)$ hardly exhibits a
 central peak. Though a faint central component is discernible, its
 intensity is significantly weaker than that of the spinwave peak, and
 the difference from the triangular case shown in Fig.4a is distinct.

 We then ascribe the origin of the broad central component observed away from the K point to the contribution of the
 $Z_2$ vortex-pair excitations. Near the K point, the central peak of
 $S(\bm{q},\omega)$ predominantly probes the dynamics of free $Z_2$
 vortices, while, as one moves away from the K point, the central
 component tends to probe the dynamics of more
 tightly bound $Z_2$-vortex pairs. Our observation that the width of the
 central component becomes significantly larger when one moves away from
 the K point (see Fig.3a and Fig.4a) means that the time scale of underlying dynamics becomes considerably faster. This seems fully consistent with our identification above, since tightly bound pairs tend to yield faster dynamics associated with pair annihilation and creation than that of free vortices. Another feature to be noticed is that, at the $\bm{q}$ points away from the K point, {\it e.g.\/}, at the M point, the central component of $S(\bm{q},\omega)$ tends to persist even at $T < T_v$. This observation also supports our interpretation above, since the vortex pair can stably exist even below $T_v$ unlike the free vortex.

\begin{figure}
 \begin{center}
 \includegraphics[]{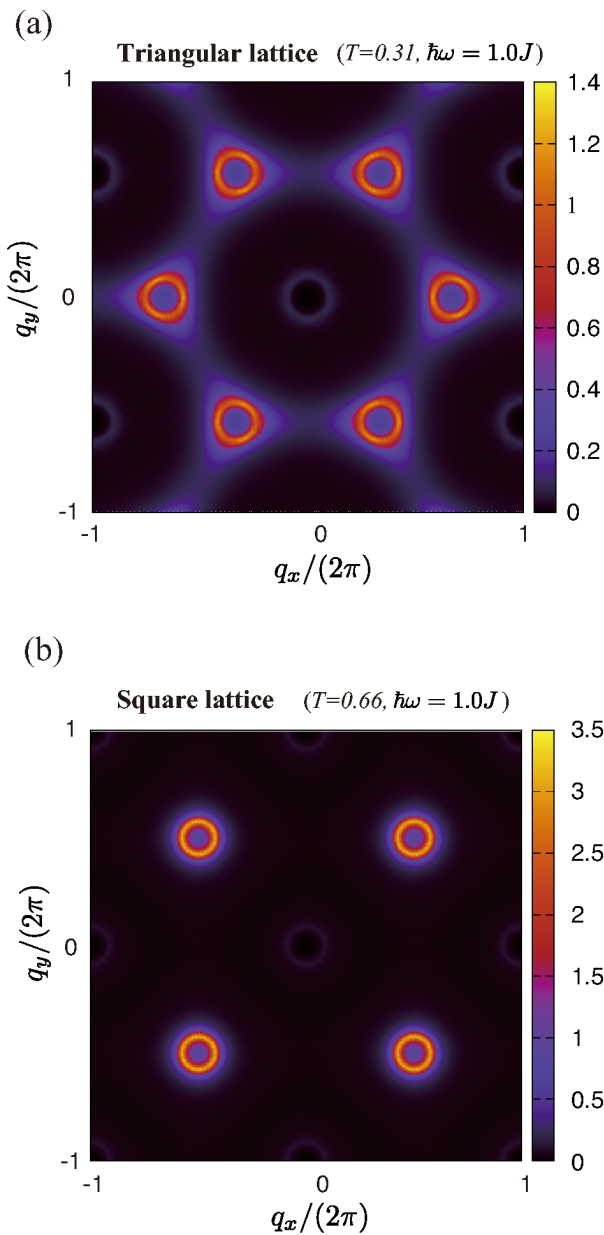}
  \end{center}
 \caption{(Color online)
(a),(b) the intensity maps of the dynamical spin structure factor in the  wavevecotr space with $\hbar \omega = J$, for the case of the triangular-lattice Heisenberg antiferromagnet with  $L=192$ at a temperature $T=0.31$ (a), and for the case of the square-lattice Heisenberg antiferromagnet with  $L=256$ at a temperature $T=0.66$ (b).}
\end{figure}

 In order to illustrate the broad central component observed for a wide
 range of wavevecotrs, we show in Fig.5a the intensity map of
 $S(\bm{q}, \omega)$ in the entire $\bm{q}$-plane at $\hbar\omega
 /J\simeq 1$ at a temperature  slightly above $T_v$, $T=0.31$. The
 effect of the central components is observed as  the intensities near
 the Brillouin zone boundaries, which appear in addition to bright
 ring-like intensities surrounding the K points corresponding to
 spinwave excitations. We note that such intensities near the zone
 boundaries cannot be seen in the corresponding intensity map of the
 square-lattice Heisenberg AF shown in Fig.5b.

\section{Summary and Discussion}
\label{Conclusion}
 From all these observations, we conclude that the central peak of the dynamical spin structure factor $S(\bm{q}, \omega)$ is originated from the $Z_2$-vortex excitation. The bases of our conclusion can be summarized as: i) Near the K point, the central peak appears only at $T > T_v$, increasing its intensity with increasing the temperature; ii) the central peak does not appear in the unfrustrated square-lattice Heisenberg AF, a model which does not sustain the $Z_2$ vortex; and iii) the central-peak structure similar to the one observed in our simulations can be reproduced by a simplified theoretical model  based on the spinwave-vortex decoupling picture.

 In contrast to the information of the standard spin correlation length
$\xi$, which is obtainable via the elastic neutron scattering
measurements, the information of the vortex correlation length $\xi_v$,
that is the length scale diverging at the vortex-transition point
$T=T_v$, is hard to obtain. In principle, however, appropriate
information about the central peak enables one to deduce $\xi_v$. On the basis
of \eqref{sqw_ana} together with the Lorenzian forms of
$S_{sw}(\bm{q})$ and $S_v(\bm{q})$,  the ratio of the central-peak
intensity $I_c(\bm{q})$ to the spinwave one $I_{sw}(\bm{q})$ near the K
point at $\bm{q}=\bm{q}_K+\delta\bm{q}$ is approximately obtained as
$I_c/I_{sw} \sim \xi_{sw}/\xi_v$. By combining it with the information of the standard spin correlation length $\xi \sim \xi_{sw}\xi_v/(\xi_{sw}+\xi_v)$, we can estimate $\xi_v$ together with $\xi_{sw}$.

 Experimentally, the dynamical spin structure factor can be measured via
 inelastic neutron scattering. As mentioned, candidate
 triangular-lattice AFs possessing a $Z_2$ vortex
 excitation might be $\mathrm{NaCrO_2}$, $\mathrm{NiGa_2S_4}$ and organic
 compounds like $\kappa$-(BEDT-TTF)$_2$Cu$_2$(CN)$_3$ or
 EtMe$_3$Sb[Pd(dmit)$_2$]$_2$. Unfortunately, the organic compounds are
 not suited to neutrons so that neutron-scattering measurements seem
 unrealistic for these materials at the present stage. Then, promising
 candidate materials for the $Z_2$-vortex detection might be
 $\mathrm{NaCrO_2}$ and $\mathrm{NiGa_2S_4}$. These materials exhibit a
 transition-like anomaly at $T\simeq 30$K ($\mathrm{NaCrO_2}$) and
 $T\simeq 8.5$K ($\mathrm{NiGa_2S_4}$), slightly below the specific-heat
 peak temperature. Often, only powder samples are available for these
 materials. Under such circumstances, we also compute the angle-averaged
 dynamical spin structure factor $S(|\bm{q}|,\omega)$ to examine how $S(|\bm{q}|,\omega)$ looks like for powder samples, and the
 result is shown in Fig.6a and 6b for both cases of the $|\bm{q}|$-value
 close to $|\bm{q}| \simeq |\bm{q}_K|$ (a), and $|\bm{q}| \simeq |\bm{q}_M|$
 (b). In both cases, the central peak is clearly observable together
 with the spinwave peaks.  Thus, inelastic neutron scattering
 experiments for powder samples would be sufficient for observing the
 signature of a $Z_2$ vortex and a $Z_2$-vortex driven topological transition. This is a nice aspect for practical purpose, since one can expect the intensity gain for powder measurements.

\begin{figure*}
 \begin{center}
  \includegraphics{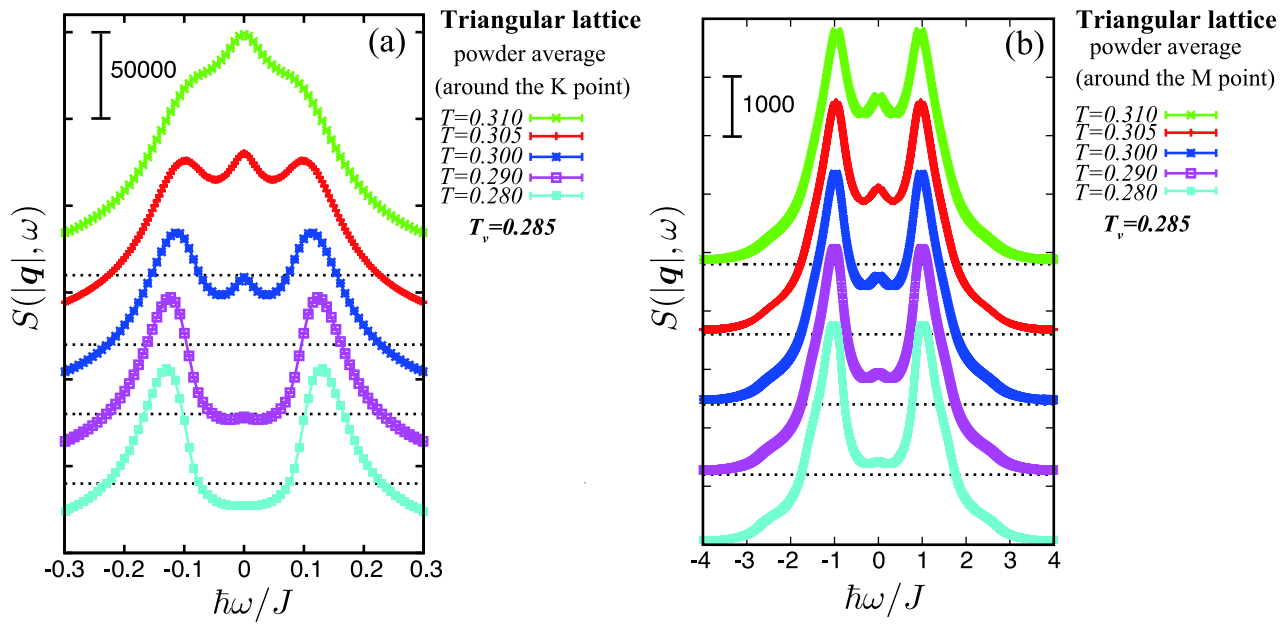}
  \end{center}
 \caption{(Color online)
 The frequency dependence of the angle-averaged dynamical spin structure
 factor of the triangular-lattice Heisenberg antiferromagnet for the
 temperatures as in Fig.3a and Fig.4a. The data should correspond to the
 neutron-scattering pattern obtained from powder samples. Panels (a) and
 (b) correspond to wavevectors satisfying $|\bm{q}_K|+4\pi/192 \leq |\bm{q}| \leq |\bm{q}_K|+6\pi/192$ (a), and
 $|\bm{q}_M| - 2\pi/384 \leq |\bm{q}| \leq  |\bm{q}_M| + 2\pi/384$ (b). }
\end{figure*}

 Inelastic neutron-scattering measurements as suggested here might provide a useful information in understanding the spin-liquid-like behavior and the transition-like anomaly observed in triangular-lattice Heisenberg AFs NaCrO$_2$ and NiGa$_2$S$_4$, and the nature of novel excitations in geometrically frustrated magnets in general.

\section*{Acknowledgements}
The authors would like to acknowledge S. Nakatsuji, M. Hagiwara,
Y. Nambu and H. Yamaguchi for useful discussion.
This work is supported by Grand-in-Aid for scientific Research on
Priority Areas ``Novel State of Matter Induced by Frustration''
(19052006). We thank the Supercomputer Center, Institute for Solid State
Physics, University of Tokyo for providing us with the CPU time.

\end{document}